\documentclass[letterpaper,10pt,twocolumn,final,oneside,conference]{IEEEtran}
\usepackage{multirow}
\usepackage{graphicx}
\usepackage{acronym}
\usepackage{amsmath}
\usepackage{amssymb}
\usepackage{booktabs}
\usepackage{eurosym}
\usepackage{siunitx}
\usepackage[caption=false,font=footnotesize,listofformat=parens]{subfig}
\usepackage{stfloats}
\usepackage{tikz}
\usepackage{hyperref}
\usepackage{multirow}
\usepackage{balance}
\usepackage{comment}

\usepackage{tikz}

\acrodef{dso}[DSO]{distribution system operator}
\acrodef{tso}[TSO]{transmission system operator}
\acrodef{ev}[EV]{electric vehicle}
\acrodef{ems}[EMS]{energy management system}
\acrodef{ls}[LS]{load shedding}
\acrodef{fls}[FLS]{fast load shedding}

\DeclareSIUnit{\kWh}{kWh}

\IEEEoverridecommandlockouts

\newcommand\copyrighttext{%
  \footnotesize
  \centering\copyright~2021 IEEE. Personal use of this material is permitted. Permission from IEEE must be obtained for all other uses, in any current or future media, including reprinting/republishing this material for advertising or promotional purposes, creating new collective works, for resale or redistribution to servers or lists, or reuse of any copyrighted component of this work in other works.\\
  IEEE EEEIC/I\&CPS Europe. DOI: \href{https://doi.org/10.1109/EEEIC/ICPSEurope51590.2021.9584758}{10.1109/EEEIC/ICPSEurope51590.2021.9584758}.}
\newcommand\copyrightnotice{%
\begin{tikzpicture}[remember picture,overlay]
\node[anchor=south,yshift=0pt] at (current page.south) {\setlength{\fboxrule}{0pt}\fbox{\parbox{\dimexpr\textwidth-\fboxsep-\fboxrule\relax}{\copyrighttext}}};
\end{tikzpicture}%
}

\begin{document}

\title{Development and Validation of a Scalable Fast Load Shedding Technique for Industrial Power Systems}

\author{%
  \IEEEauthorblockN{A. Petriccioli, S. Grillo}
  \IEEEauthorblockA{Politecnico di Milano\\Dipartimento di Elettronica, Informazione e Bioingegneria\\
  p.za Leonardo da Vinci, 32, I-20133, Milano, Italy\\
  andrea.petriccioli@mail.polimi.it, samuele.grillo@polimi.it}%
  \and
  \IEEEauthorblockN{D. Comunello, A. Cacace}
  \IEEEauthorblockA{ABB S.p.A.\\
  Process Automation -- Energy Industries\\
  via Luciano Lama, 33, I-20099, Sesto S. Giovanni, Italy\\
  \{david.comunello, andrea.cacace\}@it.abb.com}%
  }
\IEEEaftertitletext{\copyrightnotice\vspace{0.2\baselineskip}}
\maketitle

\begin{abstract}                
The work aims to improve the existing fast load shedding algorithm for industrial power system to increase performance, reliability, and scalability for future expansions. The paper illustrates the development of a scalable algorithm to compute the shedding matrix, and the test performed on a model of the electric grid of an offshore platform. From this model it is possible to study the impact on the transients of various parameters, such as spinning reserve and delay time. Subsequently, the code is converted into Structured Text and implemented on an ABB PLC. The scalability of the load shedding algorithm is thus verified, confirming its performance with respect to the computation of the shedding matrix and the usefulness of the dynamic simulations during the design phase of the plant.
\end{abstract}

\begin{IEEEkeywords}
fast load shedding, industrial power system, energy management system, offshore platform.
\end{IEEEkeywords}

\acresetall
\section{Introduction}
Industrial power systems require \acp{ems} to ensure the operations of industrial plants~\cite{Pacheco:2012,ABB}. One of the main functionalities of the \ac{ems} is the \ac{ls}. LS is used when the frequency falls below safety thresholds. If the generators are not able to stop this fall and to restore the values to the nominal ones, it is necessary to reduce the power absorption. This is where \ac{ls} comes in, taking care of shedding loads avoiding the blackout.

LS is very effective in stand-alone grids, where there are great difficulties in frequency stability related to the production of electric power, but it is also useful in grid-connected plants to handle sudden trips of the external connections. The two principal \ac{ls} methods used are the \ac{fls}~\cite{Wester:2014,Theron:2018} and the frequency-based load shedding, such as the underfrequency load shedding (UFLS)~\cite{Manson:2014}, the hybrid frequency-based load shedding~\cite{Giroletti:2012} and the neural-network underfrequency load shedding~\cite{Hsu:2011}. These systems can be integrated, and generally the UFLS is the backup for the FLS~\cite{Rajan:2019}. The FLS is by definition the fastest and it is event-driven: it requires measurements of power and circuit breaker status from the generators; its architecture is described in~\cite{Pinceti:2002}.

Dynamic simulations are very useful to optimize parameters and to check the effectiveness of the LS algorithm~\cite{Nagpal:2001,Eckl:2016,Hamilton:2009}. FLS functionalities in the industrial sector generally does not have a high degree of scalability. The algorithm proposed in this paper was developed starting from an existing offshore plant, but was structured with an intrinsic scalability that would allow its expansion and application to different plants and scenarios. Moreover, dynamic models were developed to study the impact of parameters and to optimize their choice, minimizing the load to shed to maintain stability.

The load shedding algorithm is generally implemented in the distributed control system (DCS) of the industrial plant~\cite{Mihirig:2006}, therefore one important requirement to meet was to develop the code also in Structured Text, ensuring the compatibility with PLCs, maintaining a high degree of scalability.

\section{Proposed algorithm}
\begin{figure}
  \centering
  \includegraphics[width=\columnwidth]{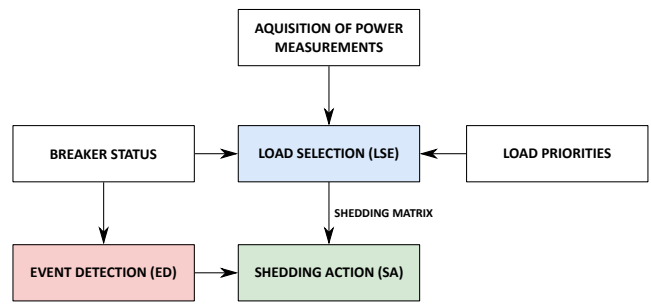}
  \caption{General structure of the algorithm.}\label{fig:struct}
\end{figure}

The structure of the FLS algorithm is described in Fig.~\ref{fig:struct}. It is important to introduce the concepts of ``event'' and ``priority'': the events are the undesired and unplanned perturbations that may bring the system to instability. For every possible event foreseen in the design process, the load shedding algorithm has to calculate the loads to disconnect in order to keep the system in safe conditions. The priority of the load represents how much the load is important for plant operations. The priority may change during the operations of the plant, so it can be set dynamically. There is the possibility that more loads have the same priority. The most important loads can be set as ``non sheddable''.\IEEEpubidadjcol

The proposed algorithm can be divided in two parts. The first one is the load selection algorithm (LSE). The inputs of this section are:
\begin{itemize}
  \item The load list with their priorities;
  \item The graph of the grid;
  \item The number of foreseen events.
\end{itemize}

This section is dived in two subsections with different purposes. The first subsection is the power mismatch (PM) calculation: here the minimum power to shed in order to maintain stability for every event is computed, according to the status of the grid. It is important to notice that the opening of one bustie does not imply a loss in the power generated, but the division in two sub-networks. For each sub-network the PM is computed, and if it is positive a shedding action is required.
The second subsection is the load selection: in this second part the algorithm selects the load to shed for every PM (thus for every event) according to their priority and to the status of the circuit breakers. The sum of the power of the shed load is the actual power shed (PS). PS must always be greater than PM. If there are loads with the same priority, the usual choice is to first shed the one with the higher power absorbed, in order to minimize the number of loads switched off.

The output of the load selection (LSE) algorithm is the shedding matrix (SM), that represents a list of loads to switch off for every event foreseen. The rows of the matrix represent the loads, while the columns represent the events. An example of matrix is shown in Table~\ref{tab:matrix}.

\begin{table}
  \centering
  \caption{Example of the shedding matrix.}\label{tab:matrix}
  \includegraphics[width=\columnwidth]{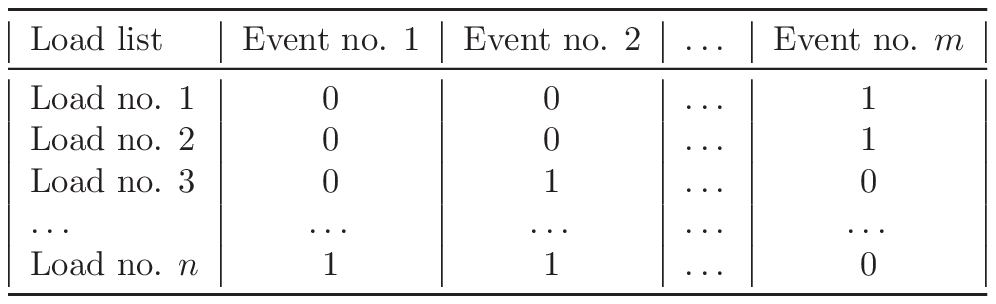}
\end{table}

The value 1 is associated to the load to be shed, while 0 to the load not to be shed.  It is useless to update the shedding matrix faster than cycle time of the acquisition of power measurements, so generally the LSE algorithm is executed every 0.5--2 seconds.
The second part of the algorithm is the event detection and shedding action (ED-SA) This second part uses as inputs:
\begin{itemize}
  \item The shedding matrix SM previously computed
  \item The status of the circuit breakers
\end{itemize}
The purpose of this section is to detect if an event occurs and take the action needed to shed the load according to the SM. When an event occurs, this section of the algorithm is triggered, and produce a signal to open the correspondent circuit breaker of the load as output. Therefore ED-SA must be executed faster, and the delay time between the event and the end of the breaking phase of the circuit breaker is generally 100 – 250 ms.
The events considered by the proposed algorithm are:
\begin{itemize}
  \item Trip of a single generator
  \item Opening of one bustie (and so reconfiguration of the grid)
  \item Loss of every generator inside a building (fire \& gas event)
  \item Blackout of the external grid
\end{itemize}

The proposed LSE algorithm works without limits on the number of generators and load, and up to three busbars. The topology of the grid is not preset, but it is an input: this ensures a high scalability. It also works when one or both the busties are open. The code is developed in MATLAB, that allows a simple management of the matrices, with many pre-set functions. From the MATLAB environment it is possible for example to import and export code in C, generate executables, perform dynamic simulations on Simulink to verify the functioning of the algorithm and from here export Structured Text (ST) code for programmable logic controllers (PLC) and other controllers. PLCs offers the advantage to be a real-time system with a defined cycle time selected during design phase. Once the ST code has been developed, it can be downloaded to different PLCs.
The algorithm considers gas turbines functionalities, e.g., gas turbines output power losses with rising temperatures and changes in the combustion mode with Dry Low Emission (DLE). In the range of the combustion change, the SR parameter is generally zero (or very low) due to gas turbine combustion instability. Generally, this range of power is avoided during normal operation, but it may occur in some situations. Thus, for safety reasons, it is considered in this study.
The algorithm allows to handle a second event: after a first event (e.g., the trip of a generator), the algorithm do not stop all the operations, but allows a second event algorithm to handle any issues while the main algorithm is restarted. This functionality can be easily implemented in a PLC, programming to stop the FLS algorithm for a few seconds (to settle the power transient) and relying on a UFLS algorithm in the meantime.

\begin{figure}
  \centering
  \includegraphics[width=\columnwidth]{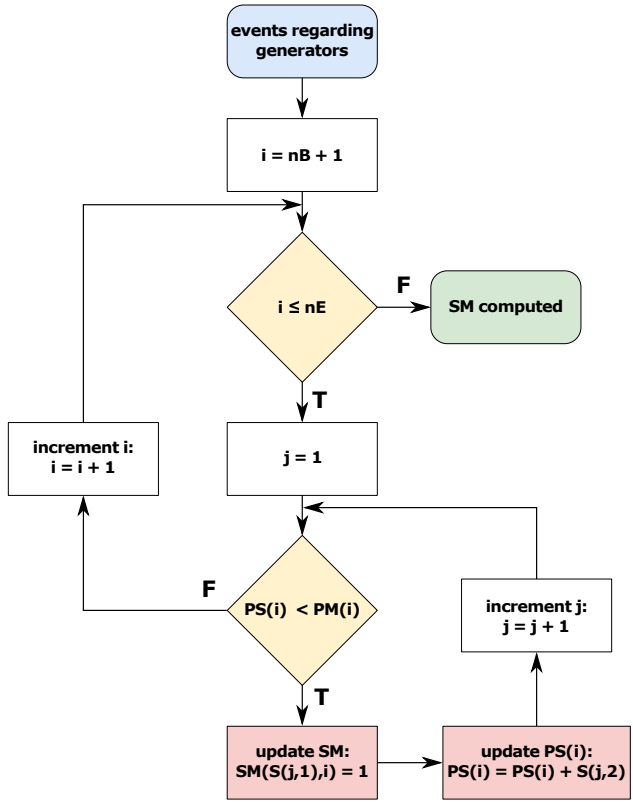}
  \caption{Flowchart for the events regarding generators.}\label{fig:flowchart}
\end{figure}

Fig.~\ref{fig:flowchart} represents the flowchart for the events regarding the generators, the other events present minimal changes with respect to the presented one. The power mismatch (PM) is computed as
\begin{equation}
  {\rm PM}_i = P_{{\rm gen}_i} - \left( {\rm SR}_{\rm tot} - {\rm SR}_i\right),
\end{equation}
where $P_{{\rm gen}_i}$ is the power output of the generator that trips, while $\left( {\rm SR}_{\rm tot} - {\rm SR}_i\right)$ is the spinning reserve of the remaining generators. The spinning reserve for this kind of application is defined as the instantaneous pickup capability, i.e., the step change in the power that a generator can take without going to underfrequency condition. Sometimes it is called incremental reserve margin (IRM). This parameter is usually given from the manufacturer of the generator, but it may be changed performing dynamic studies on the grid, as will be shown.

\section{Validation and results}
\begin{figure}
  \centering
  \includegraphics[width=\columnwidth]{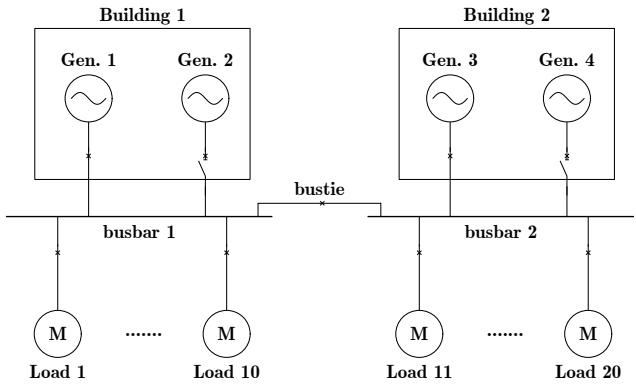}
  \caption{One-line diagram of the power system under study.}\label{fig:rete}
\end{figure}

The effectiveness of the algorithm has been tested with dynamic simulations on a model of an industrial power system. The model created represents an existing offshore platform, whose diagram is represented in Fig.~\ref{fig:rete}. Connections to the external transmission grid are not generally used: all the power is produced locally by four gas turbines.

The starting point of the dynamic model is the swing equation, which describes the rate of change of the frequency (ROCOF):
\begin{equation}
  \frac{df}{dt} = \frac{f_0}{2\sum_{i=1}^{n}H_i S_{{\rm n}_i}}\left(P_{\rm g} - P_{\rm l}\right),
  \label{eq:swing}
\end{equation}
where $H_i$ and $S_{{\rm n}_i}$ are the inertia constant and the nominal apparent power of the $i$-th generator respectively, $f_0$ is the nominal frequency, $P_{\rm g}$ is the power produced by the generators and $P_{\rm l}$ is the power absorbed by the loads. From \eqref{eq:swing} it is possible to see the variation of the frequency in the system after the event (e.g., the loss of power of one generator). Stand-alone industrial systems tend to have lower inertia and, conversely, higher per-unit accelerating power (the difference between the generated and the requested power) compared to national power systems, resulting in high frequency oscillations. The variation of the frequency is the input to the turbine-governor model, that has been tuned to represent narrowly the behavior of the gas turbines used in the plant, whose data are reported in Table~\ref{tab:param}.

\begin{table}
  \centering
  \caption{General electrical characteristics of the simulated turbine. In brackets the usual ranges for rated power and inertia constant are indicated.}\label{tab:param}
  \includegraphics[width=\columnwidth]{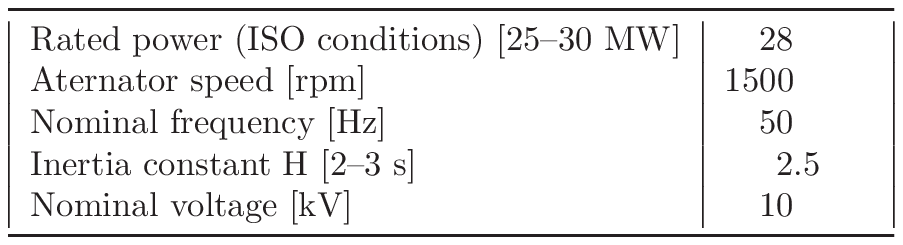}
\end{table}

The regulating power and inertia of the loads are neglected because they vary according to the loads connected and their impact is limited, moreover it is safer to neglect their effect rather than to consider them.

Simulations have been performed considering two generators connected, that is the harshest condition since the inertia is low and the accelerating power is high. The event considered at $t = \SI{2}{\second}$ is the most frequent one: the trip of a generator. The delay time include the ED-SA time, the delay in the communication and the opening of the circuit breakers of the loads; here it was assumed equal to \SI{200}{\milli\second}.

\begin{figure}
  \centering
  \includegraphics[width=\columnwidth]{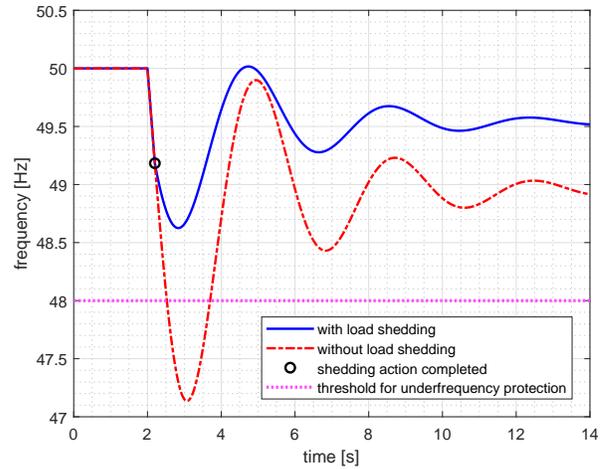}
  \caption{Frequency behavior with (blue solid line) and without (red dashed line) the shedding action.}\label{fig:freq}
\end{figure}

Fig.~\ref{fig:freq} represents the frequency with and without the shedding action. Without the LS, the nadir of frequency is \SI{47.14}{\hertz}. This value would have triggered the protection relay because it is under the underfrequency threshold, that here is set at \SI{48}{\hertz}. With the load shedding action, the nadir of frequency becomes \SI{48.63}{\hertz}, avoiding the trip of the protection devices and allowing the grid to remain in operation, even if at reduced power.

\begin{figure}
  \centering
  \includegraphics[width=\columnwidth]{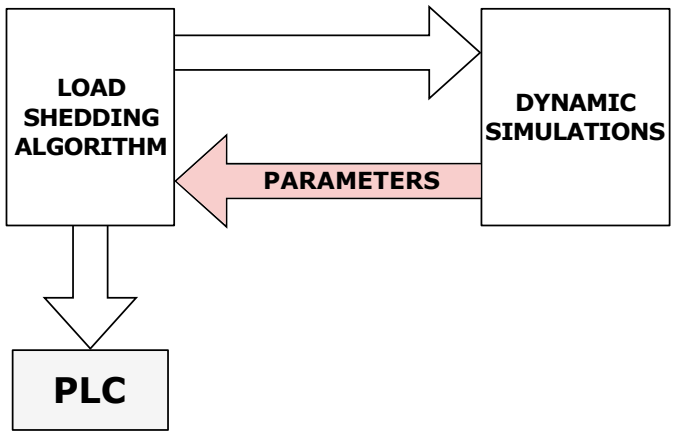}
  \caption{Graphical description of the process for the selection and optimization of the parameters.}\label{fig:proc}
\end{figure}

The dynamic model developed can be used not only for LS algorithm validation, but also to optimize the parameters. This process is represented in Fig.~\ref{fig:proc}.

Once the model of the generator is selected, its parameters of the turbine-governor model and H are set. Hence, the choice of the SR parameter used as input in the LS algorithm and total time delay remains, which must be made respecting the underfrequency specifications and the safety threshold defined.

\begin{figure}
  \centering
  \includegraphics[width=\columnwidth]{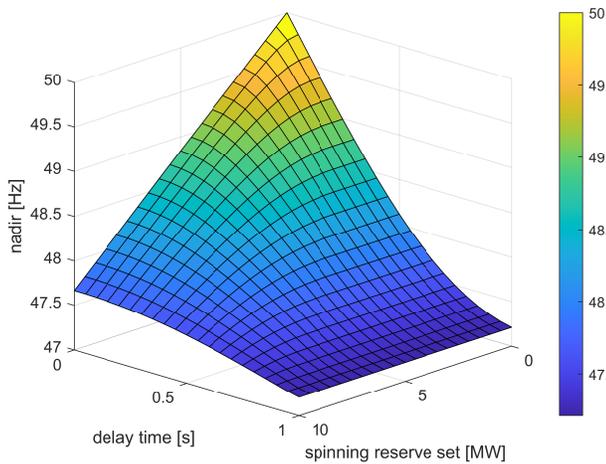}
  \caption{Frequency behavior with (blue solid line) and without (red dashed line) the shedding action.}\label{fig:surf}
\end{figure}

Performing many dynamic simulations, the trend of the nadir can be plot obtaining a surface of correlation between spinning reserve, delay time and nadir. The plot in Fig.~\ref{fig:surf} gives a visual representation of the frequency nadir as spinning reserve and delay time vary. At every intersection of delay time and spinning reserve set, a simulation has been performed, and the resulting nadir has been plotted.

Fig.~\ref{fig:surf} is very useful in the correct design of the \ac{ls} function of the plant, allowing an optimal selection of the SR parameter and the maximum delay time needed to match the design requirements, minimizing loss in operations and avoiding the trip of the underfrequency protection relays.

\begin{figure}
  \centering
  \includegraphics[width=\columnwidth]{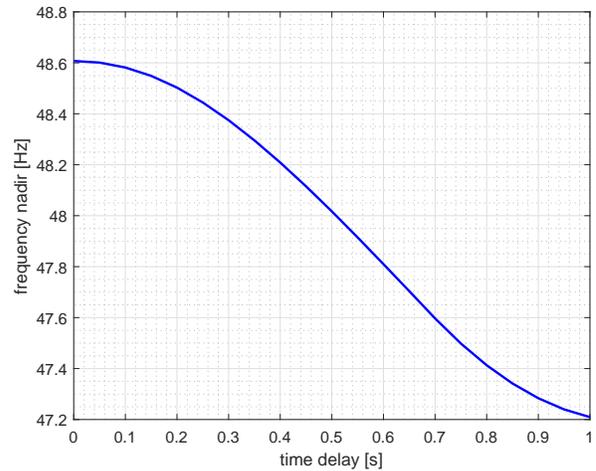}
  \caption{Trend of the frequency nadir with respect to delay time variation.}\label{fig:tdel}
\end{figure}

Once one of these two value is set, it is possible to extract a plot that highlights the impact of the other parameter. Fig.~\ref{fig:tdel} shows the impact of the delay time when SR set at \SI{6}{\mega\watt}. Up to \SI{200}{\milli\second} there is not a great difference, but the impact increases rapidly after that value.

\begin{figure}
  \centering
  \includegraphics[width=\columnwidth]{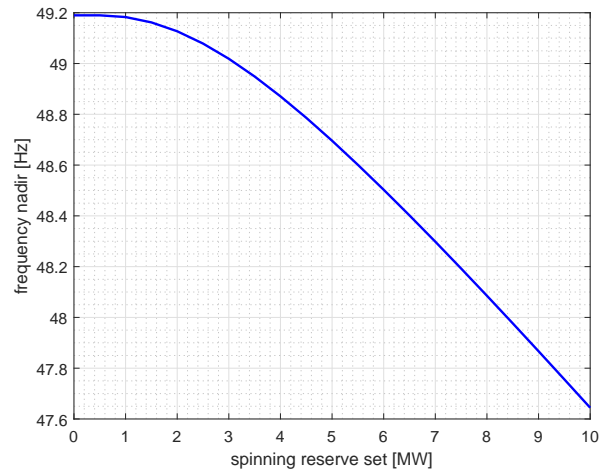}
  \caption{Trend of the frequency nadir with respect to SR variation.}\label{fig:sr}
\end{figure}

Fig.~\ref{fig:sr} shows the impact of the SR parameter when the delay time is \SI{200}{\milli\second}.

The nadir decreases because SR here is defined as the parameter that is given as input to the \ac{ls} algorithm to calculate the PM (the amount of load to shed). Considering a higher spinning reserve, the load shed would be less, and therefore the turbine will have to supply more power reaching a lower nadir.

Therefore, two different plants can select different SRs despite having the same turbine and an identical control system, according to the safety margin on the underfrequency threshold to obtain. The value of the spinning reserve set for the \ac{ls} algorithm is very important, because the higher this value is, the lower the load to shed is, allowing the plant to maintain a good degree of operation.
As long as in the plant under analysis the delay time cannot be lowered, only the SR parameter can be selected. From Fig.~\ref{fig:sr}, the maximum value of the SR to use remaining over the \SI{48}{\hertz} threshold is \SI{8.4}{\mega\watt}, however it is advisable to consider a safety margin. To obtain a safety margin of \SI{0.5}{\hertz} on the underfrequency threshold a SR = \SI{6}{\mega\watt} must be selected, while if the safety margin requested is 1 Hz this value must be lowered to \SI{3.1}{\mega\watt}.

After this optimization, the developed algorithm has been converted in Structured Text (ST) and imported in the CODESYS environment, using the automatic code generation. The ST code has been then downloaded on an ABB AC500 PLC for testing. Software tests in the CODESYS developing environment and hardware tests on the PLC were conducted to verify the computation of the shedding matrix. The ST is IEC 61131-3 compliant, ensuring the compatibility of the code with any PLC.

\section{Conclusion}
The developed FLS algorithm adapts according to the topology of the grid, thus allowing high scalability, relying only on the inputs provided from the field or defined in the design. The dynamic behavior of the algorithm was tested on a model of an existing offshore platform. From this model it was possible to study the impact on the transients of various parameters, such as spinning reserve, inertia and delay time. A model-based method was developed to obtain the optimal value of the spinning reserve to set in the algorithm, to ensure both safety and operability. The LS algorithm was coded in Structured Text, making possible to implement the algorithm on any PLC. The implementation on PLC allows to integrate this algorithm on existing control systems using the common industrial communication protocols, expanding the functionalities of the plants without the need to completely change the control system.

\section*{Acknowledgment}
The Authors would like to thank Gabriele Nani with ABB Process Automation, Genova, and Enrico Ragaini with ABB Electrification, Bergamo for their valuable comments and support. The work leading to this paper has been developed under the framework provided by the Joint Research Center between ABB and Politecnico di Milano.


\end{document}